\documentclass[aip,graphicx]{revtex4-1}
\usepackage{color}
\usepackage{amsmath}
\usepackage{mathtools}
\usepackage{url}
\usepackage{here}
\usepackage{CJKutf8}
\begin{document}

\title{Decomposition  of Friction Coefficients to Analyze Hydration Effects on a C$_{60}$(OH)$_{\rm n}$} %Title of paper

\author{Tomoya Iwashita}
\affiliation{Department of Chemistry, 
            Kyushu University, 
            Fukuoka,
            812-0395, 
            Japan}
            
\author{Yuki Uematsu}
\affiliation{Department of Physics and Information Technology, 
            Kyushu Institute of Technology,
            Iizuka, 
            820-8502,
            Japan}

\author{Masahide Terazima}
\affiliation{Department of Chemistry,
            Kyoto University, 
            Kyoto,
            606-8502, 
            Japan}

\author{Ryo Akiyama}
\email{rakiyama@chem.kyushu-univ.jp}
\affiliation{Department of Chemistry, 
            Kyushu University, 
            Fukuoka,
            812-0395, 
            Japan}

\date{\today}

\begin{abstract}
To analyze hydration effects on macromolecular diffusion, the friction coefficients of macromolecules were examined using molecular dynamics simulations with an all-atom model. In the present study, a method was introduced to decompose the molecular friction coefficient into the contributions for each site on the macromolecule. The method was applied to several fullerenols in ambient water. The friction coefficients for the hydrophilic part, such as the OH group, were larger than those for the hydrophobic part, such as the C. The hydration effect did not depend only on the kind of functional group but also on surface roughness. This approach would be useful in explaining the experimentally observed large changes in diffusion coefficients of proteins that were accompanied by conformation changes. 
\end{abstract}


\maketitle %\maketitle must follow title, authors, abstract and \pacs


\section{Introduction}
Brownian motion plays an essential role in transport phenomena. Einstein\cite{Einstein1} and Smoluchowski\cite{Smoluchowski1} studied the Brownian motion, and independently proposed the Einstein--Smoluchowski (ES) equation:
\begin{equation}
\label{ESeq}
D = \frac{ k_{\rm B} \it T }{\gamma},
\end{equation}
where $D$, $k_{\rm B}$, $T$, and, $\gamma$ are the diffusion coefficient, Boltzmann constant, temperature, and friction coefficient, respectively. For a spherical particle of radius $R$, moving in a solvent of shear viscosity $\eta$, Stokes's law predicts that the friction coefficient $\gamma$ is $6\pi \eta R$. Einstein combined the Stokes law with Eq. (\ref{ESeq}) and obtained the Stokes--Einstein (SE) law:
\begin{equation}
\label{SEeq}
D = \frac{ k_{\rm B} \it T }{6\pi \eta R},
\end{equation}
where $R$ is called the hydrodynamic radius. The SE law shows the relation between particle size and the diffusion coefficient, and has been used in particle size determination. If the SE law was exact and $R$ was the molecular size, we would not expect a small conformation change to cause a drastic change in the protein diffusion coefficient.  Terazima and coworkers have shown drastic changes occurring for various proteins, although the sizes of these proteins did not change drastically\cite{terazima1,terazima2,terazima3,terazima4,terazima5,terazima6,terazima7,terazima8}. They also discussed the drastic diffusivity changes caused by the hydration changes, for example, changes from hydrophobic solvation to hydrophilic\cite{terazima1}.\par

Nakamura et al. have studied theoretically the diffusion coefficient of a solute particle with the radial distribution functions between the solute particle and solvent particles\cite{Yoshimori1, Yoshimori2, Yoshimori3, Yoshimori4, Yoshimori5}. They adopted the hard sphere mixture model and obtained the diffusion coefficient of the solute particle. The diffusion coefficient depended on the solvation structure. Their studies showed an explicit relationship between diffusion coefficient and solvation structure. However, a detailed discussion of the experiments is problematic because proteins diffuse in an aqueous solution, and the hydration is strongly affected by the direct interaction between the water molecules and the proteins.\par

Generally, globular protein has an inhomogeneous surface comprising a mosaic patterns of hydrophilic and hydrophobic surfaces. The contributions of each surface on the diffusion behavior must depends on the surface properties because the friction between the surface and water increases the hydration becomes stronger. Thus, we could discuss the diffusion change caused by the surface modification or the small structure change, if we had a method to divide the diffusion coefficient into the contributions of specific sites on the solute surface, directly. It is, however, impossible to divide the diffusion coefficient into the various contributions. Thus, we focus on the friction coefficient in Eq. (\ref{ESeq}). The friction coefficient $\gamma$ is defined as the ratio between the drag force $F_d$ exerting  on a solute in a solvent and the velocity of the solute relative to the solvent, i. e. $\gamma=F_{d}/v$. Then, the drag force exetring on the solute molecule composed of $N$ atoms can be divided into the $N$ contributions corresponding to each atom. In principle, the friction coefficient of the molecule can be decomposed into the $N$ contributions unlike diffusion coefficient. \par

Because proteins have rough and flexible surfaces, it is difficult to analyze and discuss the contribution of each site. Before protein studies, we adopted fullerenols as an alternative to proteins. Because fullerenols also have mosaic surfaces, and the shape is not flexible, we can expect that their analysis would be simpler than that of proteins. Some researchers have investigated the diffusion coefficients of fullerenols using molecular dynamics (MD) simulations\cite{Chaban,Keshri,Maciel}. These authors found that their values decreased with the number of OH groups, and they attributed this decrease to the increased number of hydrogen bondings. Their findings  are  consistent with the arguments of Terazima and coworkers\cite{terazima1}. However, the friction coefficient has not been decomposed into the contribution for each site. Therefore, our aim is to discuss  the friction coefficient on the basis of the decomposition.\par

In the present study, we examine a method to decompose the friction coefficient into the contributions for specific sites on a polyatomic molecule using MD simulations. The decomposition method is formulated in the framework of the generalized Langevin equation\cite{Mori1}. The MD simulations were conducted with a fullerene or a fullerenol in an aqueous solution and applied the method to the solute. Thus, the dependence of the friction coefficient on the number of OH groups was discussed based on the analysis.\par

\section{Theory and Methods}
\subsection{Definitions of contribution functions and values}
Mori introduced the generalized Langevin equation (GLE) based on a projection operator method\cite{Mori1}. The dynamics of Brownian motion can be described by the GLE. The one-dimensional GLE for the Brownian particle (solute) is given by
\begin{equation}
\label{GLEeq}
F(t) = -\int_{0}^{t} d\tau \Gamma(\tau) v(\tau - t) + F_{\rm R}(t),
\end{equation}
where $F$, $\Gamma$, $v$, and $F_{\rm R}$ are total force, friction kernel, velocity of the mass center relative to a solvent, and a random force, respectively. Here, the convolution integral represents the drag force, which takes into account the memory effect.  We can obtain the friction coefficient by integrating the friction kernel:
\begin{equation}
\gamma = \int_{0}^{\infty} d\tau \Gamma(\tau).
\end{equation}
The random force $F_{\rm R}$ is uncorrelated with the initial velocity:
\begin{equation}
\label{v0FReq}
\langle v(0) \cdot F_{\rm R} (t)  \rangle = 0,
\end{equation}
and is connected to the friction kernel by the fluctuation-dissipation theorem\cite{Kubo1}:
\begin{equation}
\label{FDTeq}
\langle F_{\rm R} (0) \cdot F_{\rm R} (t)  \rangle = k_{\rm B}T \Gamma(t).
\end{equation}
\par

We now discuss the decomposition of the friction coefficient. In the solute molecule composed of $N$ atoms, the random force $F_{\rm R}$ can be described as the sum of the random forces exerting on the $i$-th atom,  $f_{\rm R}^i(t)$, as follows:
\begin{equation}
\label{frdefeq}
F_{\rm R}(t) = \sum_{i=1}^{N} f_{\rm R}^i(t).
\end{equation}
Thus, the fluctuation-dissipation theorem (Eq. (\ref{FDTeq})) can be rewritten as:
\begin{equation}
\label{Kdefeq}
\Gamma(t) = \sum_{i=1}^{N}\bigg\lbrack \frac{1}{k_{\rm B}T} \langle f_{\rm R}^i(0) \cdot F_{\rm R}(t) \rangle\bigg\rbrack = \sum_{i=1}^{N} K^i(t),
\end{equation}
where $K^i(t)$ is defined as $1/k_{\rm B}T \langle f_{\rm R}^i(0) \cdot F_{\rm R}(t) \rangle$. Here, we can regard the $K^{i}(t)$  as the contribution of the $i$-th atom to the friction kernel. In the present paper, we refer to $K^i(t)$ as a contribution kernel of the $i$-th atom. The GLE (Eq. (\ref{GLEeq})) can be rewritten as:
\begin{equation}
\label{GLEatomeq}
F(t) = \sum_{i=1}^{N}\bigg\lbrack - \int_{0}^{t} d\tau K^i(\tau) v(\tau -t) + f_{\rm R}^i(t) + f_{\rm IN}^i({\mathbf q}(t))\bigg\rbrack,
\end{equation}
where $f^i(t)$ is the total force, $f_{\rm IN}^i({\mathbf q}(t))$ is an intramolecular force exerting on $i$-th atom. The intramolecular forces are determined by the solute conformation ${\mathbf q}(t)$ at that time $t$, and are defined by the coordinate derivative of the intramolecular potential $V_{\rm sol}({\mathbf q}(t))$ as follows:
\begin{equation}
\label{findefeq}
  f_{\rm IN}^i({\mathbf q}(t))=-\frac{\partial V_{\rm sol}(\mathbf q(t))}{\partial q^i},
\end{equation}
where $q^i$ is a coordinate of the $i$-th atom. The summation of the intramolecular forces is zero, i.e. $\sum_{i=1}^{N}  f_{\rm IN}^i({\mathbf q}(t))$ = 0. For example, in the all-atom MD simulations, the spring potentials, such as bonds, angles, and dihedrals, are parts of the intramolecular potential. From, Eq.(\ref{GLEatomeq}), the total force on the i-th atom, $f^i(t)$ is defined as:
\begin{equation}
\label{fidefeq}
 f^i(t)= - \int_{0}^{t} d\tau K^i(\tau) v(\tau -t) + f_{\rm R}^i(t) + f_{\rm IN}^i({\mathbf q}(t)).
\end{equation}
Then, the integrated value of the contribution kernels can also be regarded as the contribution of each atom to the friction coefficient, which is represented as:
\begin{equation}
\label{kdefeq}
\kappa^i = \int_{0}^{\infty} d\tau K^i(\tau),
\end{equation} 
where $\kappa^i$ is the contribution value of the $i$-th atom to the friction coefficient.\par

In Eq. (\ref{GLEatomeq}), $F(t)$ and $f^i(t)$ can be easily obtained from MD trajectories as forces of the potential derivatives. However, decomposing these forces into random-and-drag components requires certain techniques. Moreover, as will be discussed later, obtaining the intramolecular forces from MD trajectories is difficult. Thus, we first define the extramolecular force:
\begin{equation}
\label{fEXdefeq}
 f_{\rm EX}^i(t) = f^i(t) - f_{\rm IN}^i({\mathbf q}(t))=- \int_{0}^{t} d\tau K^i(\tau) v(\tau -t) + f_{\rm R}^i(t),
\end{equation} 
and derive an equation to extract the contribution kernel from the extramolecular forces. The treatment of the intramolecular forces will be addressed subsequently (See Section \ref{sec:details}).\par

To extract the contribution kernel, the $f_{\rm EX}^i(0)$ is multiplied by Eq. (\ref{GLEeq}) to take the ensemble average, i.e.:
\begin{equation}
\label{Kdef2eq0}
\langle f_{\rm EX}^i(0) \cdot F(t)\rangle =  -\int_{0}^{t} d \tau \Gamma(\tau) \langle  f_{\rm EX}^i(0)\cdot v(\tau -t) \rangle + \langle f_{\rm R}^i(0) \cdot F_{\rm R}(t)\rangle,
\end{equation} 
and apply Eq. (\ref{Kdefeq}), which leads to the following equation: 
\begin{equation}
\label{Kdef2eq}
K^i(t) = \frac{1}{k_{\rm B} T}\bigg\lbrack  C^{f(\rm EX,i) F}(t) + \int_{0}^{t} d \tau \Gamma(\tau) C^{f(\rm EX,i) v}(t-\tau) \bigg\rbrack
\end{equation}, 
where $C^{f(\rm EX,i) v}(t)=\langle f_{\rm EX}^i(0) \cdot v(t)\rangle$, $C^{f(\rm EX,i) F}(t)=\langle f_{\rm EX}^i(0) \cdot F(t)\rangle$. If we have $C^{f(\rm EX,i) v}(t)$, $C^{f(\rm EX,i) F}(t)$, and $\Gamma(\tau)$, the contribution kernels can be calculated using Eq. (\ref{Kdef2eq}).\par

\subsection{MD simulation setup}
\begin{figure}[htb]
	\centering
	\includegraphics[width=\linewidth]{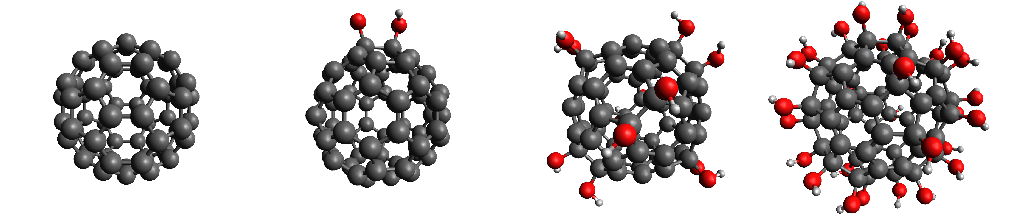}
	\caption{The structures of fullerene and fullerenols. From left to right, C$_{60}$, C$_{60}$(OH)$_{2}$, C$_{60}$(OH)$_{12}$, and C$_{60}$(OH)$_{28}$.}
	\label{fig:str}
\end{figure}

MD simulations were conducted to obtain the diffusion properties including the friction coefficient. The simulation system was composed of 1000 SPC/E  water \cite{SPCE} molecules and a solute in a cubic box with the periodic boundary condition. We also performed MD simulations of systems with the number ratios of solute-to-water molecules 1/1500 and 1/2200 to confirm the system size effect \cite{YHM}.  Fullerene and fullerenols were adopted as the solute molecules. The structures are shown in Fig. \ref{fig:str}. The force-field Generalized Amber Force Field 2 (GAFF2) was used for the solute molecules. The solute parameters were generated by AmberTool\cite{Amber} with the AM1--BCC charge model\cite{AM1BCC}. Before the parameterization, the structure was optimized by the HF/6-31G* level using Gaussian 16\cite{g16}. Cases in which all the solute charge parameters were set to zero were also investigated. In this study, these solutes are referred to as ``nonpolar''.\par

The production runs were performed under $NVT$ conditions. The ambient conditions, $T =$ 298.15 K and $P =$ 1 bar ( = $10^5$ Pa),  were adopted. The temperature was controlled by the stochastic velocity rescaling thermostat \cite{V_rescale} coupled with a time constant of 0.5 ps to the solvent only. This thermostat does not affect the solute friction kernel\cite{Daldrop1}. Each simulation cell size, $L$, as well as the volume $L^3$, were determined by the average cell size under the $NPT$ condition at a pressure of $P =$ 1 bar, which was controlled by a stochastic cell rescaling barostat\cite{C_rescale}. After equilibration, the production simulations were conducted under more than 40 independent trajectories for a total of 1 $\sim$ 3$\mu$s. We performed all MD simulations using the GROMACS 2021.4 software\cite{GROMACS} with particle mesh Ewald summation (PME) \cite{PME}. The cutoff of 1.2 nm was used for the direct space of PME and the non-bonded van der Waals interactions. All the solute bonds were kept fixed using the LINCS algorithm\cite{Hess1}, and the bonds and angles of the water molecule were kept fixed by the SETTLE algorithm\cite{Settle}. The equations of motion were integrated using the leap-frog algorithm with a time-step of 2 fs.\par

\subsection{Computational details of the decomposition of friction coefficients}
\label{sec:details}
 To obtain $C^{f(\rm EX,i) v}(t)$ and $C^{f(\rm EX,i) F}(t)$, we recall the intramolecular forces in Eq.(\ref{fEXdefeq}). The intramolecular force $f_{\rm IN}^i$ can be divided into two main components, namely, the intramolecular force $f_{\rm gas}^i$, which is defined in the vacuum, and the force $f_{\rm solv}^i$, caused by the solvation. Because the solute structure deformation changes the solvation-free energy, the solvent affects the solute potential, and $f_{\rm solv}^i$ is a kind of mean force that includes the shielding effect and surface tension. In MD simulations, $f_{\rm gas}^i$ can be easily calculated from the solute conformation , such as such as bond, angle, and dihedral, according to the force field, which was GAFF2 in the present study. The direct calculation of $f_{\rm solv}^i$ is difficult, and it prevents us from obtaining the intramolecular and extramolecular forces directly. Thus, we introduce a way of calculating $C^{f(\rm EX,i) v}(t)$ and $C^{f(\rm EX,i) F}(t)$ without the direct calculation of $f_{\rm EX}^i$ or $f_{\rm IN}^i$.\par

If we have another MD trajectory with the same solute conformations ${\mathbf q}_0$, which is the solute conformations ${\mathbf q}(0)$ in the original MD trajectory, and different solvent configurations, $C^{f(\rm EX,i) v}(t)$ can be calculated as follows:
\begin{equation}
\label{eq_anov}
\begin{split}
C^{f(\rm EX,i) v}(t) 
&=\langle f^i_{\rm EX}(0) \cdot v(t) \rangle \\
&=\langle f^i_{\rm EX}(0) \cdot v(t) \rangle - \langle f^i_{\rm EX,ano} \cdot v(t) \rangle \\
&=\langle \lbrack f^i_{\rm EX}(0) - f^i_{\rm EX,ano} \rbrack \cdot v(t) \rangle \\
&=\langle \lbrack f^i_{\rm EX}(0) - f^i_{\rm EX,ano} + f^i_{\rm IN}({\mathbf q}_0) - f^i_{\rm IN,ano}({\mathbf q}_0)\rbrack \cdot v(t) \rangle \\
&=\langle \lbrack f^i_{\rm EX}(0)  + f^i_{\rm IN}({\mathbf q}(0))- \lbrace f^i_{\rm EX,ano} + f^i_{\rm IN,ano}({\mathbf q}_0) \rbrace \rbrack \cdot v(t) \rangle \\
&= \langle\lbrack f^i(0)-f^i_{\rm ano}\rbrack \cdot v(t) \rangle,
\end{split}
\end{equation} 
where $f^i_{\rm ano}$, $f^i_{\rm EX,ano}$, and $f^i_{\rm IN,ano}({\mathbf q}_0)$ are the total, extramolecular, and intramolecular forces exerting on the $i$ -th atom in another MD trajectory. Here, $f^i_{\rm IN}({\mathbf q}_0) - f^i_{\rm IN,ano}({\mathbf q}_0)$ = 0 because the solute conformation and solvent type are the same, and $\langle f^i_{\rm EX,ano} \cdot v(t) \rangle = 0$, as long as the solvent flow is independent of that of the original MD trajectory. In addition, $C^{f(\rm EX,i) F}(t)$, can be calculated as follows:
\begin{equation}
\label{eq_anof}
C^{f(\rm EX,i) F}(t) = \langle\lbrack f^i(0)-f^i_{\rm ano}\rbrack \cdot F(t) \rangle.
\end{equation} 
Therefore, $C^{f(\rm EX,i) v}(t)$ and $C^{f(\rm EX,i) F}(t)$ can be calculated using $f^i-f^i_{\rm ano}$ instead of $f_{\rm EX}^i=f^i-f^i_{\rm IN}$.\par

In this study, to obtain the $f^i_{\rm ano}$ for ${\mathbf q}_0$, i.e. $f^i_{\rm ano}({\mathbf q}_0)$, additional simulations were conducted in which the solute conformation was frozen at the target solute conformation ${\mathbf q}_0$, whereas the water molecules moved freely. Here, the correction for the velocity of the center of mass was turned off, which means that the center of a solute molecule can move freely relative to the solvent molecules. Such an approach is called a belly approach \cite{belly}. An MD simulation with the belly approach is called a belly simulation in this paper. In a belly simulation, each solute atom is assumed to be trapped in an infinitely strong bias potential $W({\mathbf q})$, which is a function of the solute conformation $\mathbf q$ with the following conditions: 
\begin{equation}
\label{bias}
W({\mathbf q}_0)=0,\  {\rm and} \ -\frac{\partial}{\partial {\rm q}^i} W({\mathbf q}_0)=0,
\end{equation}
where ${\rm q}^i$ is a coordinate of the $i$ -th solute atom. In constrain simulations, the relationship of the mean force $f_{\rm solv}^i({\mathbf q})$ and the mean force in constrain simulations $f_{\rm solv,res}^i({\mathbf q})$ is:
\begin{equation}
f_{\rm solv}^i({\mathbf q})=f_{\rm solv,res}^i({\mathbf q}) - \frac{\partial}{\partial {\rm q}^i} W({\mathbf q}),
\end{equation}
where $f_{\rm solv}^i({\mathbf q}_0)=f_{\rm solv,res}^i({\mathbf q}_0)$ using Eq. (\ref{bias}). Hence, we can obtain the $f^i_{\rm ano}$ from the belly simulation without changing the term of the intramolecular force $f^i_{\rm IN}({\mathbf q}_0)$.\par

One belly simulation provides one $f^i_{\rm ano}({\mathbf q}_0)$. To take the average of the $C^{f(\rm EX,i) v}(t)$ and $C^{f(\rm EX,i) F}(t)$, we need many runs of belly simulations to obtain restart points of  $f^i_{\rm ano}({\mathbf q}_0)$. Then, the equilibration steps for each belly simulation require a high computational cost. Thus, we joined the belly simulations to each other using several quasistatic steps. The details are described in the supplementary informations (See section. 1).\par

For the extraction of the friction kernel $\Gamma (t)$, the iterative method developed by Kowalik et al. \cite{Kowalik1} was used. The friction kernel can be obtained using a numerical method \cite{Berne2} and Laplace transform methods \cite{Allen1, Manonov1} using the first kind Volterra integral equation: 
\begin{equation}
\label{V1eq}
m\dot{C}^{vv}(t)=m \langle v (0) \cdot F (t)  \rangle = -\int_{0}^{t} d\tau \Gamma(\tau) C^{vv}(\tau - t),
\end{equation}
which is obtained from the GLE by multiplying Eq. (\ref{GLEeq}) by the initial velocity $v(0)$ and taking an ensemble average \cite{Berne1, Shin1}. However, these calculations lead to the stability problems \cite{Lange1}. To improve the algorithms, Kowalik et al. integrated Eq. (\ref{V1eq}) and derived the following equation:  
\begin{equation}
\label{Geq}
mC^{vv}(t)= -\int_{0}^{t} d\tau G(\tau) C^{vv}(\tau - t) + mC^{vv}(0),
\end{equation}
where $G(t)$ is running integral over the friction kernel defined as
\begin{equation}
\label{Gdefeq}
G(t)= \int_{0}^{t} d\tau \Gamma(\tau).
\end{equation}
They showed that the friction kernel is obtained stably by first calculating $G(t)$ from Eq. (\ref{Geq}) and then obtaining the slope of $G(t)$\cite{Kowalik1}. The iterative method is represented by the following  discretized equation with time-step $\Delta t$ and frame-number j = 0, 1, 2, ... ,
\begin{equation}
\label{Gdeq}
G_j= \frac{2}{\Delta t C_0^{vv}} \bigg\lbrack mC_0^{vv} - mC_j^{vv} - \sum_{k=0}^{j-1} \Delta t G_j C_{j-k}^{vv} w_{j,k} \bigg\rbrack,
\end{equation}
where $G_j=G(j \Delta  t)$, $C_j^{vv}=C^{vv}(j \Delta t)$, $m$ is the mass of solute and $w_{j,k}$ is a weighting factor of the trapezoidal rule defined as $w_{j,j}=w_{j,0}=1/2$ and $w_{j,k} = 1$ otherwise. After obtaining the $G(t)$ from Eq. (\ref{Gdeq}), we obtain the friction kernel $\Gamma (t)$ as follows:
\begin{equation}
\label{kobtain}
 \Gamma_j=
 \begin{dcases}
    \frac{G_1}{\Delta t} & (j = 0) \\
    \frac{G_{j+1}-G_{j-1} }{2\Delta t}  & (j \ge 1)
 \end{dcases},
\end{equation}
where $\Gamma_j$ = $\Gamma(t)$.\par

The contribution kernels $K^i(t)$ can be calculated with the obtained $C^{f(\rm EX,i) v}(t)$, $C^{f(\rm EX,i) F}(t)$ and $\Gamma (t)$ from the discretized version of Eq. (\ref{Kdef2eq}):
\begin{equation}
\label{Kdef2deq}
K^i_j = \frac{1}{mC^{vv}_0}\bigg\lbrack  C^{f(\rm EX,i) F}_j + \sum_{k=0}^{j-1} \Delta t \Gamma_j C^{f(\rm EX,i) v}_{j-k} w_{j,k}\bigg\rbrack,
\end{equation} 
where $K^i_j=K(j \Delta  t)$, $C^{f(\rm EX,i) F}_j=C^{f(\rm EX,i) F}(j \Delta  t)$, and $C^{f(\rm EX,i) v}_j=C^{f(\rm EX,i) v}(j \Delta  t)$. Here, we use the equipartition theorem: $k_{\rm B}T=mC^{vv}_0$. Then, the contribution values $\kappa^i$ can be obtained using Eq. (\ref{kdefeq}). Here, the contribution function of a specific site, such as an OH group, can be obtained directly from Eq. (\ref{Kdef2deq}) by replacing $f_{\rm EX}^i$ with the force exerting on the center of mass of that site.\par

Because the diffusion and friction coefficients depend on the simulation cell size, we applied the correction proposed by Yeh and Hummer\cite{YHM}. They show that the diffusion coefficient of an infinite system $D_0$ can be predicted by the following equation:
\begin{equation}
\label{YHeq}
D_0=D_{\rm pbc} + \frac{k_{\rm B}T\xi}{6\pi \eta L} - \frac{2 k_{\rm B}T R^2}{9 \eta L^3},
\end{equation}
where $\xi$=2.837297, $L$ is the side length of the cubic basic cell, and $D_{\rm pbc}$ is the diffusion coefficient for the finite system under periodic boundary conditions. For large $L$ values, the second order correction term $2 k_{\rm B}T R^2/9 \eta L^3$ can be ignored, but if the size of L is insufficient, the influence may be non-negligible\cite{Iwashita1}. In this study, we checked the results for different cell sizes with water: solute molecule ratios of 1:1000, 1:1500, and 1:2200. After that, we applied the second order correction with the hydrodynamic radius obtained from the $D_0$ of the ratio = 1:2200, and accepted the averaged value for the three $D_0$s with different ratios. After obtaining $D_0$ and $D_{\rm pbc}$, we scaled the contribution values $\kappa^i$ as follows:
\begin{equation}
\label{kdef0eq}
 \kappa^i_0 = \frac{D_{\rm pbc}}{D_0} \kappa^i
\end{equation} 
to obtain the contribution values of the infinite system $\kappa^i_0$.
\par

In this study, the diffusion coefficient $D_{\rm pbc}$ and shear viscosity $\eta$ were calculated using the Green--Kubo relation\cite{viscosity1,viscosity2}:
\begin{equation}
\label{GKDeq}
D_{\rm pbc} = \frac{k_{\rm B} T}{mC^{vv}(0)}\int_0^{\infty} d \tau C^{vv}(\tau),
\end{equation}
and
\begin{equation}
\label{GKetaeq}
\eta = \frac{V}{k_{\rm B} T} \frac{1}{5} \sum_i^5 \int_0^{\infty} d \tau \langle P_i(0) P_i(\tau)\rangle,
\end{equation}
where $V$ is the volume of the system and $P_i$ represents each of five independent components of the traceless stress tensor, $P_{i=1\sim5}=[(P_{xx}-P_{yy})/2,(P_{yy}-P_{zz})/2,P_{xy},P_{yz},P_{zx}]$. In Eq. (\ref{GKDeq}), the value was scaled by $k_{\rm B}T/mC^{vv}(0)$ to unify with the ideal temperature $T$. The correlation functions, such as $C^{vv}$, $C^{f(\rm EX,i) v}$ and $C^{f(\rm EX,i) F}$, are averaged over all independent directions: x, y, z.\par

\section{Results and Discussion}
The contribution  values of C$_{60}$(OH)$_{2}$, C$_{60}$(OH)$_{12}$, and C$_{60}$(OH)$_{28}$ in water were calculated. To determine the effects of electrostatic interactions, the values of nonpolarized solutes, in which the charge parameters of the solute were set to zero, were also obtained.  The latter solutes are named nonpolar solutes, such as nonpolar C$_{60}$(OH)$_{2}$. To emphasize the differences between them, the former solutes are occasionally called polar solutes, such as polar C$_{60}$(OH)$_{2}$.\par

\begin{figure}[H]
	\centering
	\includegraphics[width=0.65\linewidth]{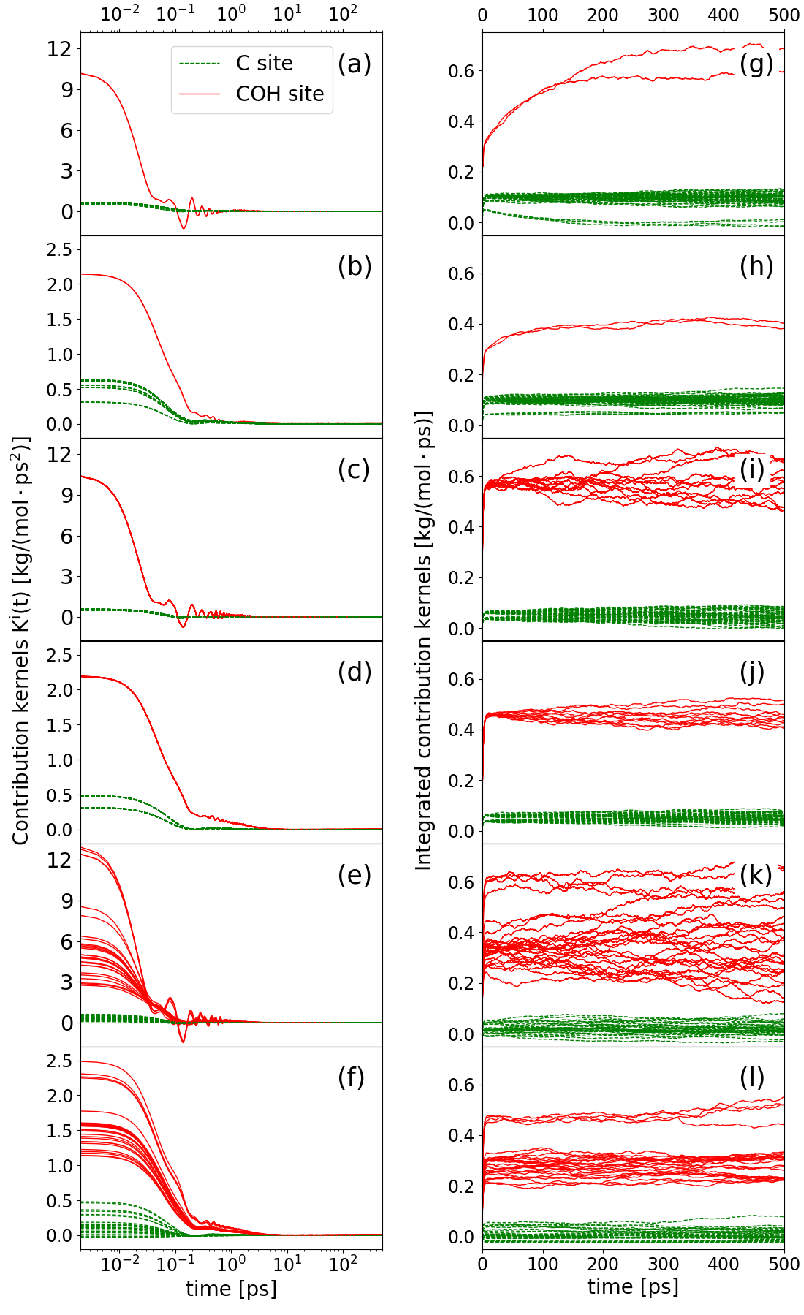}
	\caption{The contribution kernels (left panels) and the integrated contribution kernels (right panels) for 60 C sites or the COH (carbon atom and OH group) site in fullerenols. From top to bottom: C$_{60}$(OH)$_{2}$ [(a) and (g)], nonpolar C$_{60}$(OH)$_{2}$ [(b) and (h)], C$_{60}$(OH)$_{12}$ [(c) and (i)], nonpolar C$_{60}$(OH)$_{12}$ [(d) and (j)], C$_{60}$(OH)$_{28}$ [(e) and (k)] and nonpolar C$_{60}$(OH)$_{28}$ [(f) and (l)]. The red, solid, and green dashed curves show the results of the COH and C site, respectively.}
	\label{fig:cks}
\end{figure}

The contribution kernels were obtained by decomposing the friction kernels based on Eq. (\ref{Kdef2deq}), where the friction kernels were calculated using Eqs. (\ref{Gdeq}) and (\ref{kobtain}) from $C^{vv}(t)$s \cite{Kowalik1}. Fig. \ref{fig:cks} presents the contribution kernels (left panels) and the integrated contribution kernels (right panels) for each solute, (C$_{60}$(OH)$_{2}$, nonpolar C$_{60}$(OH)$_{2}$, C$_{60}$(OH)$_{12}$, nonpolar C$_{60}$(OH)$_{12}$, C$_{60}$(OH)$_{28}$, nonpolar C$_{60}$(OH)$_{28}$). The red and green solid curves show the results of the COH (a carbon atom and the OH group) and the C site, respectively.\par

\begin{figure}[H] 
	\centering
	\includegraphics[width=0.7\linewidth]{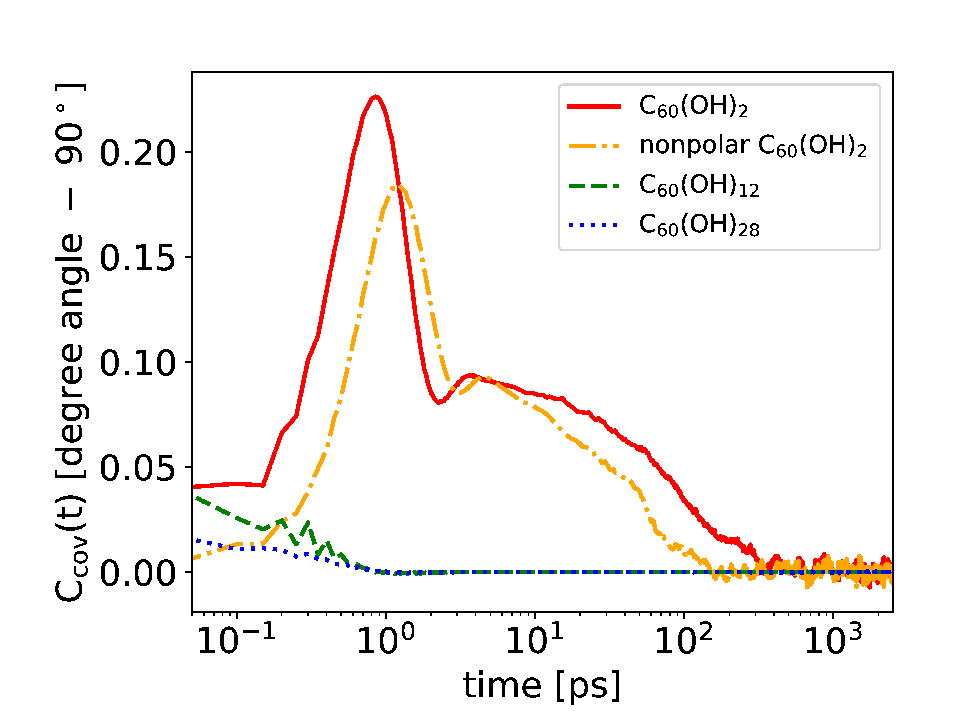}
	\caption{The angular correlation function between the velocity and direction of the  molecule defined by Eq. (\ref{Ccoveq}) where the results are converted to the degree angle and 90 degrees are subtracted. The red solid, orange dash-dotted and green dashed, blue dotted curves are the results for C$_{60}$(OH)$_{2}$, nonpolar C$_{60}$(OH)$_{2}$, C$_{60}$(OH)$_{12}$, and C$_{60}$(OH)$_{28}$, respectively. }
	\label{fig:Ccov}
\end{figure}

In Fig. \ref{fig:cks} (left panels), all of the contribution kernels converge toward zero. In Fig. \ref{fig:cks} (right panels), the value of each integrated contribution kernel reaches a plateau, which corresponds to the contribution value of each site. This behavior is consistent with the contribution kernels. With increasing structure asymmetry, the relaxation times of the contribution kernels become longer. In polar and nonpolar C$_{60}$(OH)$_2$, the integrated contribution functions reach plateau values around 300 ps, while the other functions reach values around 10 ps, which is almost the same as the relaxation time of the friction kernel of C$_{60}$. This slowing seems to be caused by the higher friction of the hemisphere containing COH sites compared to the opposite hemisphere. Consequently, a large drag force on one side and rotation in a constant direction seem to produce a flow around the molecule. If this assumption is adequate, longer relaxation times (caused by the flow) can be expected in the contribution functions. This discussion is consistent with the results of the angular correlation function between the velocity and direction of a molecule defined by
\begin{equation}
\label{Ccoveq}
C_{\rm cov}(t) = \frac{1}{n_{\rm co}}\sum_{i=1}^{n_{\rm co}}\left \langle  \frac{\mathbf{v}(0)}{\| \mathbf{v}(0) \|} \cdot \frac{\mathbf{r^i_{\rm co}}(t)}{\| \mathbf{r^i_{\rm co}}(t) \|} \right \rangle,
\end{equation}
where $\mathbf{r^i_{\rm co}}$ is the $i$--th bond vector between carbon and oxygen, $n_{\rm co}$ is the number of bonds between carbon and oxygen, and $\| \cdots \|$ denotes the norm. The results for C$_{60}$(OH)$_{2}$ (red), nonpolar C$_{60}$(OH)$_{2}$ (orange),  C$_{60}$(OH)$_{12}$ (green) and C$_{60}$(OH)$_{28}$ (blue) are shown in Fig. \ref{fig:Ccov} where the values are transformed to the degree angle and 90 degrees are subtracted because 90 degrees mean no correlation.  The functions of C$_{60}$(OH)$_{12}$ and C$_{60}$(OH)$_{28}$ quickly converge to zero, but the functions of the polar and nonpolar C$_{60}$(OH)$_{2}$ converge to zero around 300 and 200ps, respectively.\par

\begin{figure}[H]
	\centering
	\includegraphics[width=0.8\linewidth]{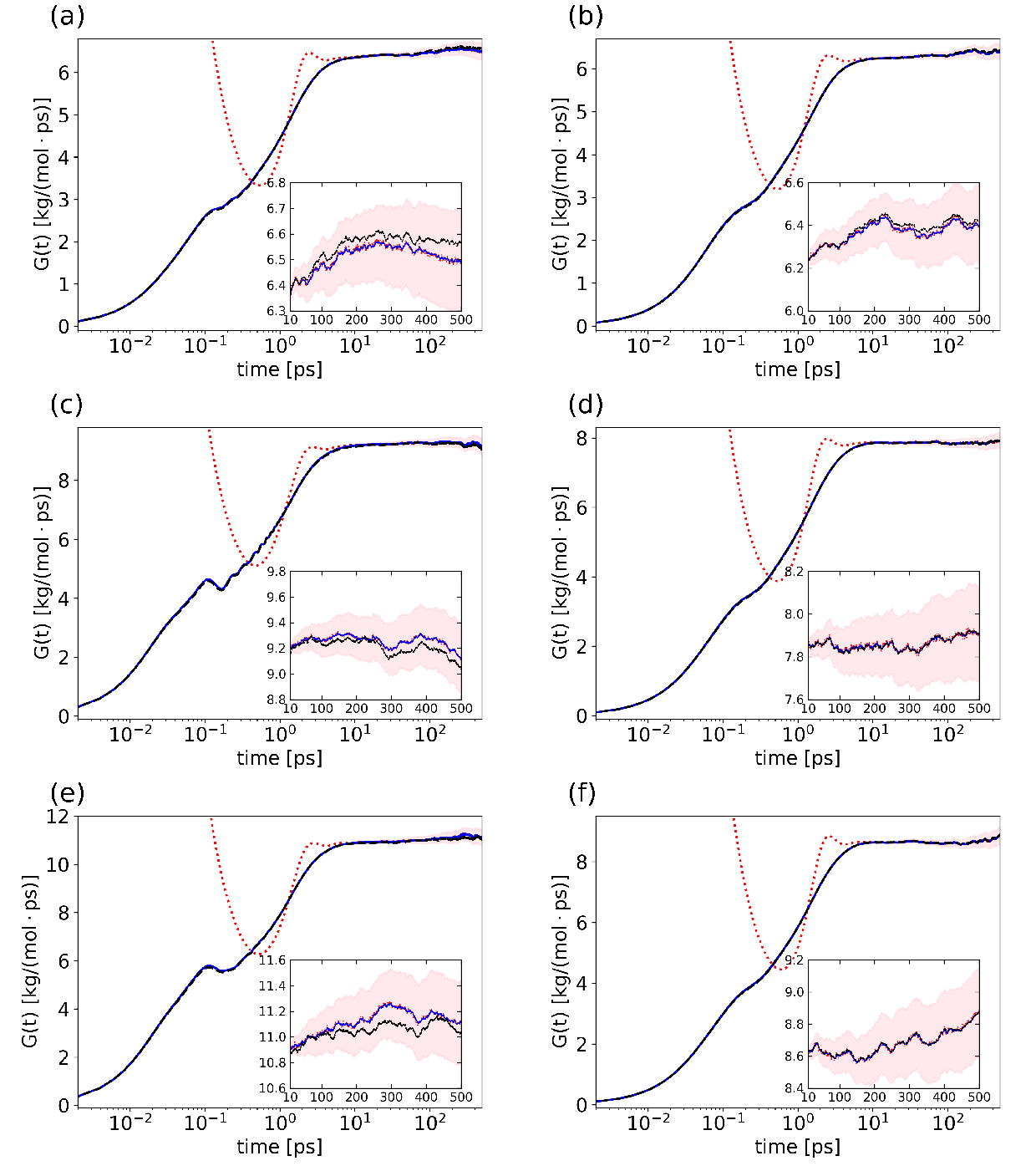}
	\caption{The summation of integrated contribution kernels (black dashed lines), the integrated friction kernels (blue solid lines) and $k_{\rm B}T/\int C^{vv}(\tau) d\tau$ (red dotted lines) for (a) C$_{60}$(OH)$_{2}$, (b) nonpolar C$_{60}$(OH)$_{2}$, (c) C$_{60}$(OH)$_{12}$, (d) nonpolar C$_{60}$(OH)$_{12}$, (e) C$_{60}$(OH)$_{28}$, and (f) nonpolar C$_{60}$(OH)$_{28}$. The pink areas denote the 95\% confidence interval for the integrated friction kernels. In the insets, the range between 10 and 500 ps is enlarged.}
	\label{fig:cktest}
\end{figure}

As indicated in Eq. (\ref{Kdefeq}), the summation of contribution kernels must coincide with the friction kernel. This requirement also extends to the total integrated contribution kernels and the integrated friction kernels. Fig. \ref{fig:cktest} shows the total integrated contribution kernel (black dashed line), the integrated friction kernel (blue solid line), and $k_{\rm B}T/\int C^{vv}(\tau) d\tau$ (red solid line) for each solute. Although the results obtained from the integrated contribution kernels slightly deviate from the integrated friction kernels, these deviations remain within the numerical errors. These results ensure that the total contribution value coincides with the friction coefficient within the numerical errors. Meanwhile, the integrated friction kernel and the total integrated contribution kernel agree with the result of $k_{\rm B}T/\int C^{vv}(\tau) d\tau$ at large $t$. Here, $k_{\rm B}T/\int C^{vv}(\tau) d\tau$ approaches $\gamma=k_{\rm B}T/D$, i.e., the ES equation (Eq. \ref{ESeq}). Thus, the friction coefficients and the total contribution values follow the ES equation.\par

\begin{figure}[H]
	\centering
	\includegraphics[width=0.8\linewidth]{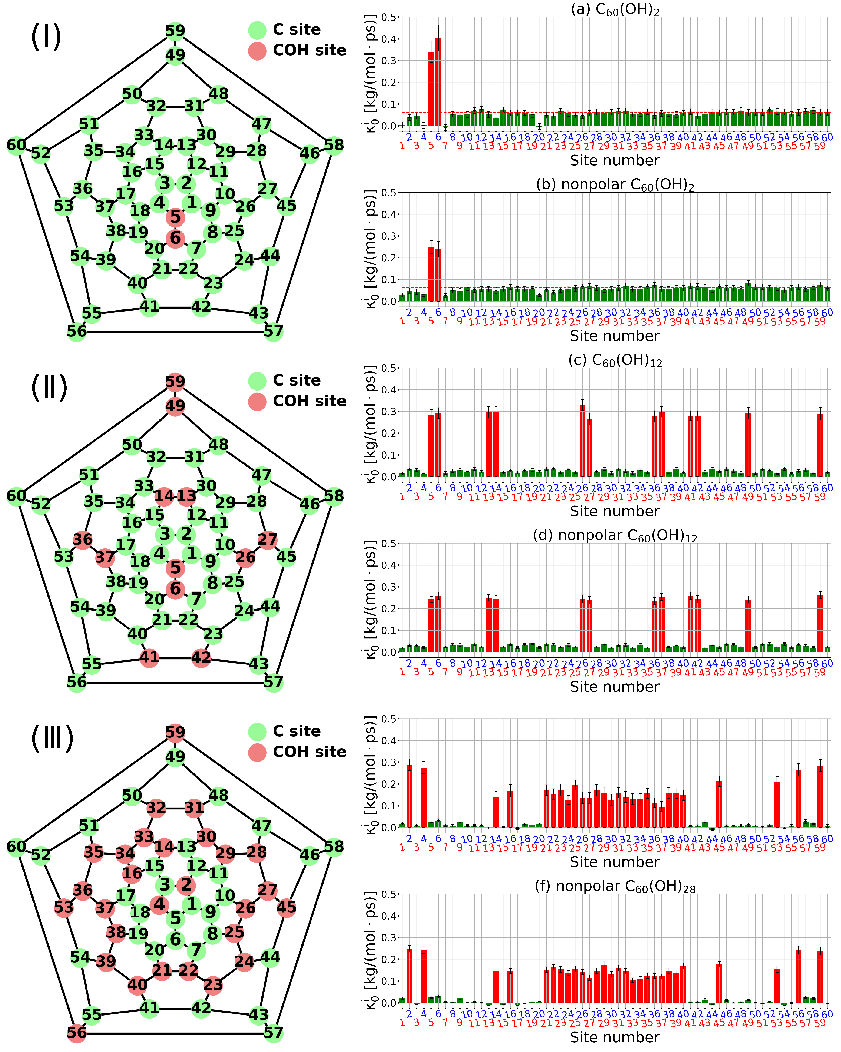}
	\caption{The contribution values of 60 (C or COH) sites in (a) C$_{60}$(OH)$_{2}$, (b) nonpolar C$_{60}$(OH)$_{2}$, (c) C$_{60}$(OH)$_{12}$, (d) nonpolar C$_{60}$(OH)$_{12}$, (e) C$_{60}$(OH)$_{28}$, and (f) nonpolar C$_{60}$(OH)$_{28}$. The red and green bars indicate the contribution values of the COH and C site, respectively. The error bars denote the 95\% confidence interval. The site numbers correspond to the numbers on the fullerene expanded maps in the left panels, where the green and red circles indicate the C and COH site, respectively. In the left panels, from top to bottom: (I) C$_{60}$(OH)$_{2}$, (II) C$_{60}$(OH)$_{12}$, and (III) C$_{60}$(OH)$_{28}$. In (a) and (b), the red dashed line indicates the friction coefficient per one C atom on fullerene (obtained by dividing the friction coefficient of fullerene by 60).}
	\label{fig:cvs}
\end{figure}

Here, the contribution values obtained from integrated contribution kernels and scaled with Eq.(\ref{kdef0eq}) are shown. Fig. \ref{fig:cvs} (right panels) shows the contribution values of 60 C or COH sites for each solute. The red and green bars indicate the contribution values of the COH and C site, respectively. The site numbers correspond to the numbers on the fullerene expanded maps in Fig. \ref{fig:cvs} (left panels), where the green and red circles indicate the C and the COH site, respectively. \par

In the case of C$_{60}$(OH)$_{2}$, the contribution values for C sites, which apart from the COH sites, are almost the same with each other in Fig. \ref{fig:cvs}(a), and similar to those for fullerene C$_{60}$. The values are almost the same as the friction coefficient per one C atom on fullerene obtained by dividing the friction coefficient of fullerene by 60 (See red dashed line in Fig. \ref{fig:cvs}(a)). The same argument applies in the nonpolarized C$_{60}$(OH)$_{2}$ (See red dashed line in Fig. \ref{fig:cvs}(b)). These results indicate that the C site contributions were correctly taken into account. By contrast, the contribution values of the COH sites and the adjacent C sites deviate from the constant value. The contribution values of the COH sites are much larger than those of C for C$_{60}$. By contrast, those of adjacent C sites are smaller than those of C for C$_{60}$.\par

\begin{figure}[hbt]
	\centering
	\includegraphics[width=0.5\linewidth]{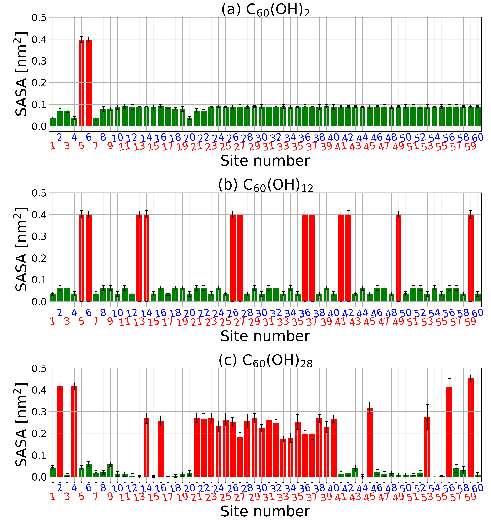}
	\caption{The solvent-accessible surface area (SASA) of 60 sites in (a) C$_{60}$(OH)$_{2}$, (b) C$_{60}$(OH)$_{12}$, and (c) C$_{60}$(OH)$_{28}$. The red and green bars indicate the SASA of the COH and C site, respectively. The error bars denote the 95\% confidence interval. The site numbers correspond to the numbers on the fullerene expanded maps in the Fig. \ref{fig:cvs} (left panels).}
	\label{fig:sasa}
\end{figure}

The difference is discussed based on the solvent-accessible surface area (SASA). Fig. \ref{fig:sasa}(a) shows the SASA of 60 sites in C$_{60}$OH$_{2}$, where each SASA was calculated using the GROMACS software (gmx sasa) \cite{GROMACS, sasa} with the Bondai's van der Waals radii\cite{bondi1} and a 0.14 nm probe radius. The figures suggest a strong correlation between the friction and the SASA. As the SASA increases, the friction increases.\par

\begin{figure}[hbt]
	\centering
	\includegraphics[width=0.45\linewidth]{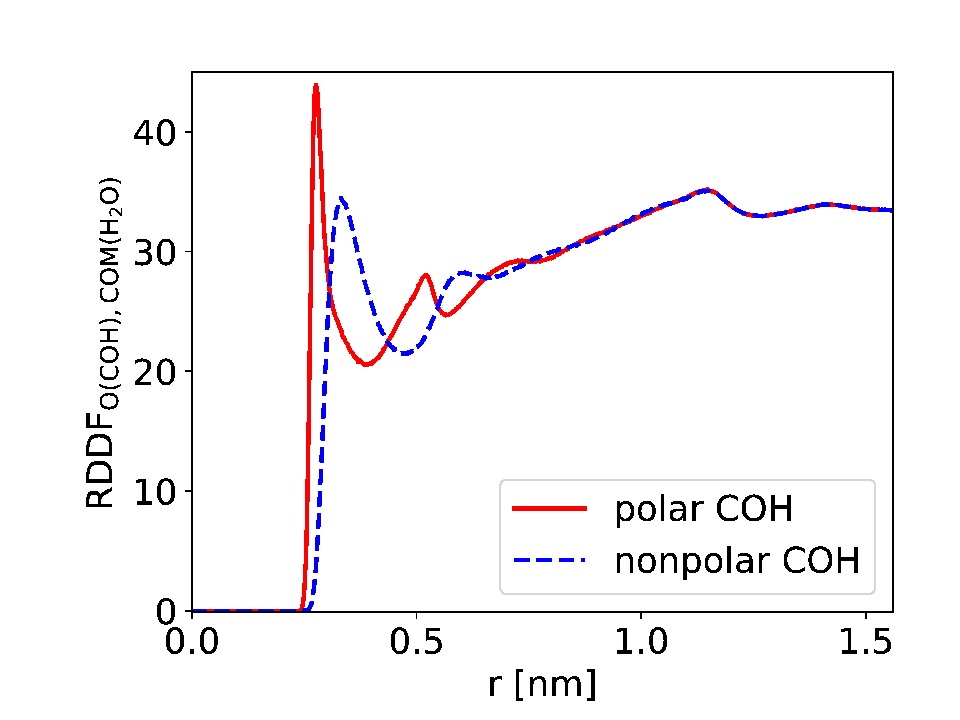}
	\caption{The radial density distribution function between the center of oxygen atom in the COH site and the mass center of water for polar (red solid line) and nonpolar (blue dashed line) C$_{60}$(OH)$_{2}$, respectively.}
	\label{fig:rddfocom}
\end{figure}

\begin{figure}[hbt]
	\centering
	\includegraphics[width=0.9\linewidth]{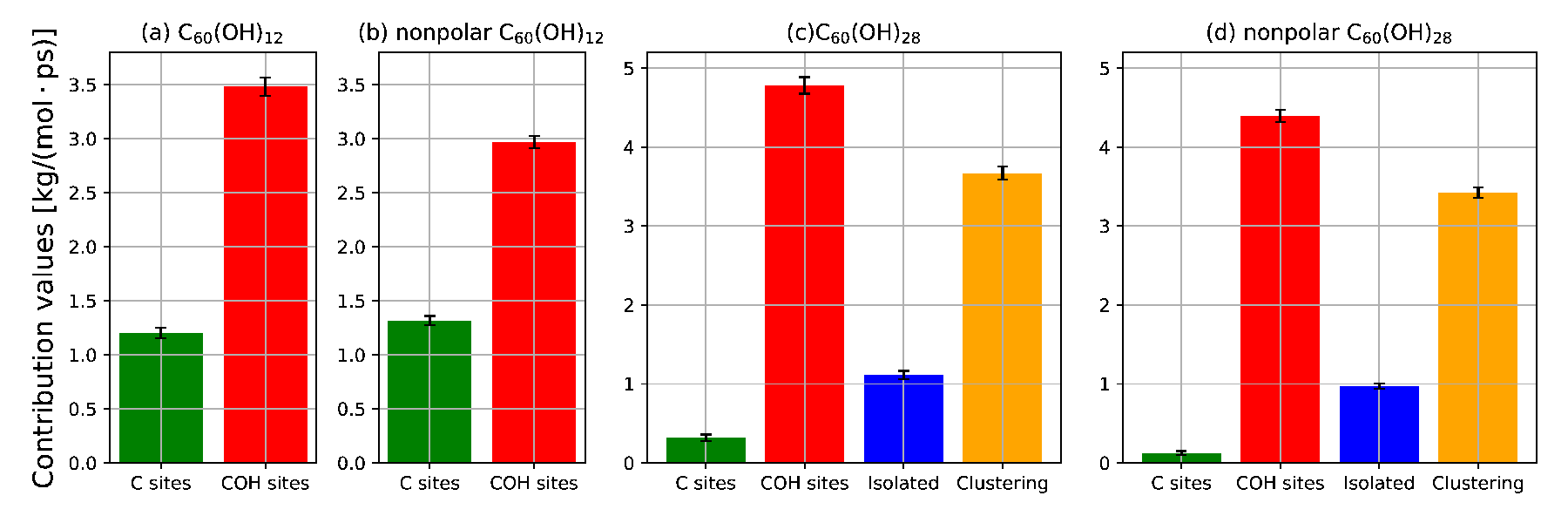}
	\caption{The total contribution values for specific sites in (a) C$_{60}$(OH)$_{12}$, (b) nonpolar C$_{60}$(OH)$_{12}$, (c) C$_{60}$(OH)$_{28}$ and (d) nonpolar C$_{60}$(OH)$_{28}$. The green, red, blue, and orange bars indicate the values for C sites, COH sites, isolated COH sites and clustering COH sites. The error bars denote the 95\% confidence interval. }
	\label{fig:cvssum}
\end{figure}

\begin{figure}[hbt]
	\centering
	\includegraphics[width=0.45\linewidth]{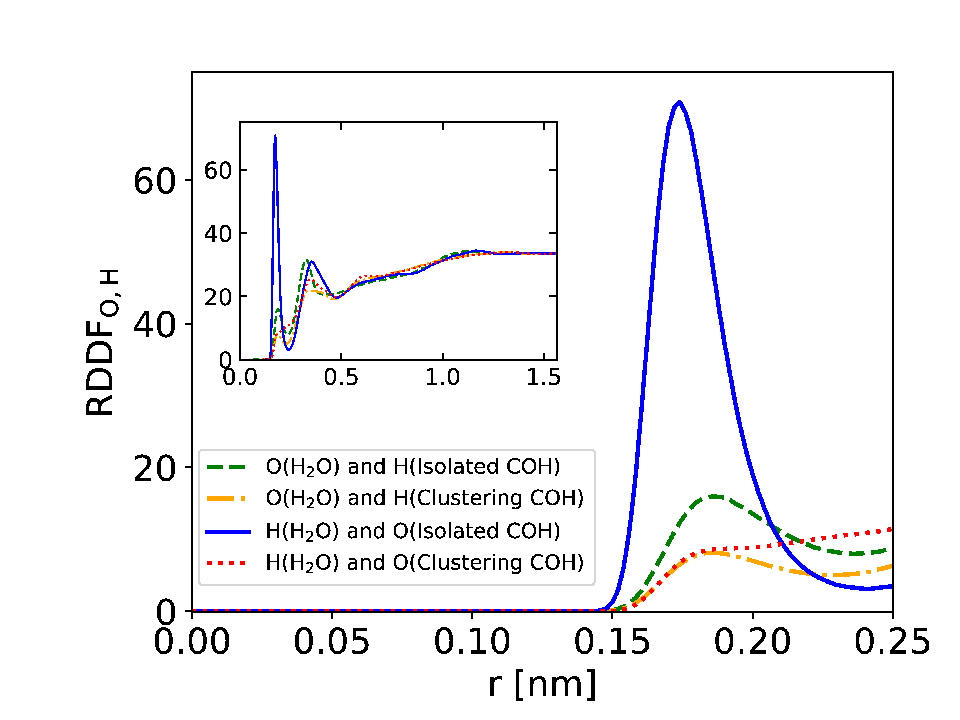}
	\caption{The radial density distribution functions (RDDF) between the oxygen of the COH site and hydrogen of water for the clustering COH sites (red dotted line) and the isolated COH sites (blue solid line), as well as the RDDF between hydrogen of the COH site and oxygen of water for the clustering COH sites (orange dash-dotted line) and the isolated COH sites (green dashed line). The RDDFs in the range $r = 0 \sim 0.25$ nm are mainly shown.  In the inset, the entire RDDF range is shown.  }
	\label{fig:rdfohho}
\end{figure}

The contribution values cannot be explained based only on the SASA. For example, in the case of C$_{60}$(OH)$_{2}$, the contribution values of COH site are larger by a factor of about 1.5 than those of the COH sites in the nonpolarized one. However, it was expected that the SASA of the COH sites would be smaller than those of the nonpolar COH ones because of the hydrogen bond between charged sites. Fig. \ref{fig:rddfocom} shows the radial density distribution function between the center of oxygen atom in the COH site and the mass center of water for polar (red solid line) and nonpolar (blue solid line) C$_{60}$(OH)$_{2}$, respectively, which indicates that water molecules can be closer to the polar COH sites than to the nonpolar ones. Therefore, the differences in contribution values of the COH sites cannot be explained only by the surface areas. As the charge on the site increases, the friction becomes stronger due to the hydrogen bond.\par 

Similar differences in the contribution values of the COH sites were also observed in the case of C$_{60}$(OH)$_{12}$ and C$_{60}$(OH)$_{28}$. To clarity the difference, we sum up the total contribution values for C and COH sites for each solute (C$_{60}$(OH)$_{12}$, nonpolarized C$_{60}$(OH)$_{12}$, C$_{60}$(OH)$_{28}$, and nonpolarized C$_{60}$(OH)$_{28}$). The results for the C sites (green bar) and the COH sites (red bar) are shown in Fig. \ref{fig:cvssum}. The contribution values of the COH sites are larger than those of the nonpolarized COH sites even in the case of C$_{60}$(OH)$_{12}$ and C$_{60}$(OH)$_{28}$. These results also imply the influence of hydrogen bonds due to the charge on the OH sites.\par

To support the above discussion of hydrogen bonds, the COH sites on the C$_{60}$(OH)$_{28}$ were divided into two groups: clustering (site=14,16,21,22,23,24,25,26,27,28,29,30,31,32,33,34,\\35,36,37,38,39,40,45,53) or isolated (site=2,4,56,59) COH group, where the site numbers correspond to the numbers on the fullerene expanded map (See Fig. \ref{fig:cvs}).\par

In Fig. \ref{fig:cvssum}(c) and (d), the total contribution values for the clustering and isolated COH sites in the polarized and nonpolarized C$_{60}$(OH)$_{28}$ were also plotted. The results again show the influence of the hydrogen bonds. Let us compare the contribution values on the isolated sites (blue bars). The value for the polar C$_{60}$(OH)$_{28}$ was larger than that for the nonpolar C$_{60}$(OH)$_{28}$. A similar difference was observed in the results for the clustering sites (orange bars). However, the increases per site for the dense COH site (1.0 $\times$ 10$^{-2}$ kg/[mol$\cdot$ps] ) is much smaller than the value for the isolated COH site (3.5 $\times$ 10$^{-2}$ kg/[mol$\cdot$ps] ) or the value for the COH site in C$_{60}$(OH)$_{12}$ (4.3 $\times$ 10$^{-2}$ kg/[mol$\cdot$ps]). \par

The difference between the clustering and isolated sites is discussed based on the hydration structure. Fig.\ref{fig:rdfohho} shows the RDDF between the COH site and the water molecule. The RDDFs between the hydrogen of the COH site and the oxygen of water for the clustering COH sites and for the isolated COH sites are plotted in orange and green solid curves, respectively. The RDDFs between the oxygen of the COH site and the hydrogen of water for the clustering COH sites and for the isolated COH sites are plotted in red  and blue solid curves, respectively. The first peak's height is almost proportional to the probability of hydrogen bond between the solvent water and the COH sites, i.e., the macromolecular surface.\par

In Fig. \ref{fig:rdfohho}, the first peak for the clustering COH sites (red and orange curves) is lower than the peaks for the isolated one (blue and green curves). This difference in the first peak is significant in the RDDF between the hydrogen of the COH sites and the water's oxygen. This result suggests that the reduction in accessible space for the solvent water molecules and the formation of internal hydrogen bonds among the COH sites in the clustering COH sites reduce the number of hydrogen bonds between water and COH in the clustering COH sites.\par

Chaban and Fileti also studied the hydration structures around fullerenols and discuss the internal hydrogen bonds\cite{Chaban}. Our analysis supports their arguments. In increasing the contribution value for the friction, the isolated COH sites are more effective than the clustering COH sites because of the hydrogen bonds between them and solvent water. These results indicate that the number of hydrogen bonds has an influence on the contribution value for friction and support the argument of Terazima and coworkers on the reduction of the diffusion coefficient due to the hydrogen bond formation between a protein and water molecules \cite{terazima1,terazima2,terazima3,terazima4,terazima5,terazima6,terazima7,terazima8}.\par

\begin{figure}[hbt]
	\centering
	\includegraphics[width=0.8\linewidth]{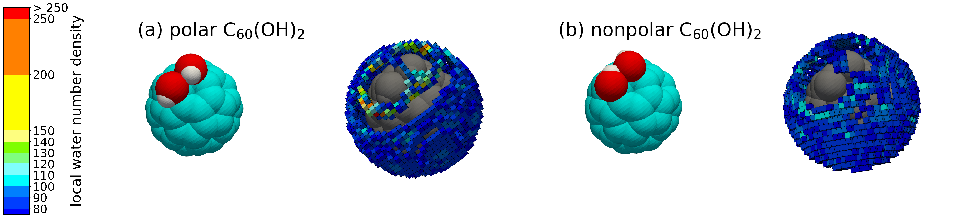}
	\caption{The spatial density distribution functions (SDDF) of water around (a) polar and (b) nonpolar C$_{60}$(OH)$_2$ are depicted using a 0.05 nm cubic grid resolution. In (a) and (b), the images on the left show the camera angles and the images on the right show the SDDF grids with the local water number density larger than 75. Each grid is classified by the local water number density, and colored according to the left color bar.}
	\label{fig:sddf_oh2}
\end{figure}

In the above paragraph, we showed that COH sites increase the contribution values of the sites. However, the existence of a COH site affects the contributions of other sites, and particularly the neighboring sites of the COH site. For example, the contribution values of C sites (site=1,4,7,20) near the COH site decrease in the case of C$_{60}$(OH)$_2$ (See Fig. \ref{fig:cvs} (a)). When the COH sites became nonpolar, the contribution values of C sites (site=1,4,7,20) increase slightly (See Fig. \ref{fig:cvs} (b)). In both cases, it can be argued that the SASA reduction induced by the COH sites causes the value reduction based on the above discussion on the SASA and friction. Moreover, it can be expected that the difference between the polar COH and the nonpolar COH is caused by the difference in hydration strength at the COH site.\par

To discuss the hydration strength at the COH site, we obtain the spatial density distribution function (SDDF) of water molecules around C$_{60}$(OH)$_{2}$. Fig. \ref{fig:sddf_oh2} shows the SDDF around (a) polar or (b) nonpolar C$_{60}$(OH)$_{2}$ using 0.05 nm cubic grid resolution, where the grids with the local water number density larger than 75 are visualized. The grids surrounding the neck region of the COH site show high water density in nonpolar C$_{60}$(OH)$_{2}$, whereas in polar C$_{60}$(OH)$_{2}$, such high-density grids are absent. By contrast, the grids around the hydrogen-bonding sites at the polar OH groups are high-density vales. These high-density values mean a high probability of water molecule existence. The water molecule near the polar OH groups reduces the accessibility of nearby C sites for water molecules. As a result, the contribution values of these neighboring C sites in polar C$_{60}$(OH)$_{2}$ become lower than those near nonpolar C$_{60}$(OH)$_{2}$.\par

\begin{figure}[hbt]
	\centering
	\includegraphics[width=\linewidth]{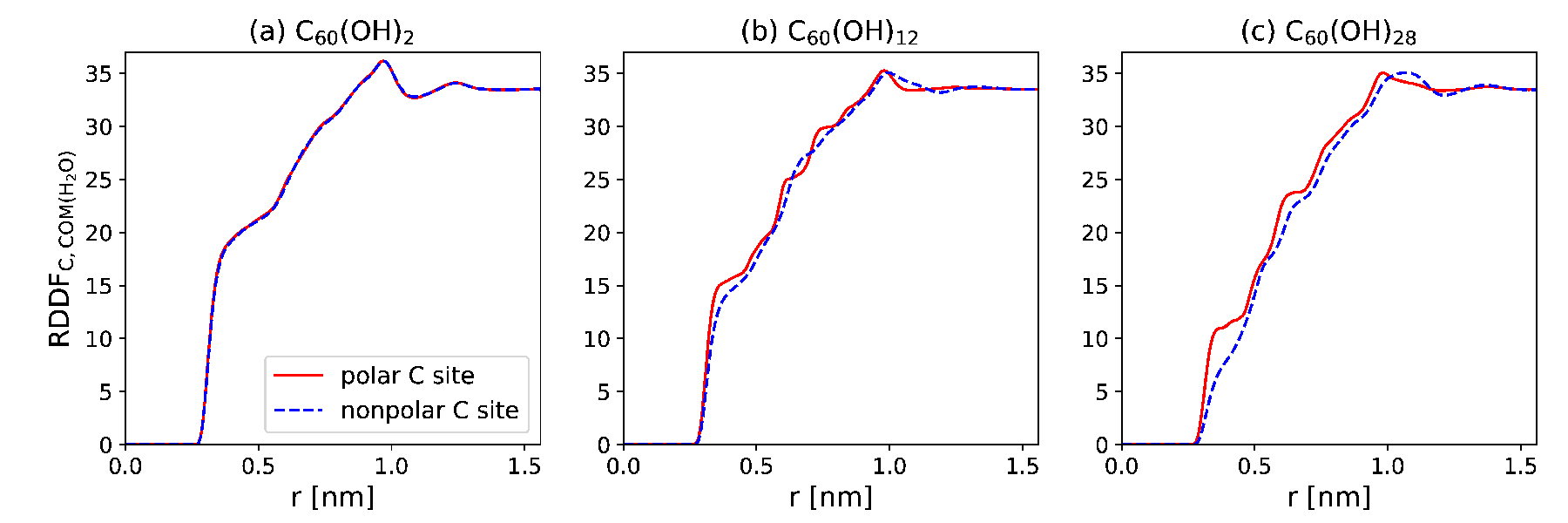}
	\caption{The radial density distribution functions (RDDF) between the C site and the mass center of water in both the polar (red solid line) and nonpolar (blue dashed line) cases for (a) C$_{60}$(OH)$_{2}$, (b) C$_{60}$(OH)$_{12}$ and (c) C$_{60}$(OH)$_{28}$.}
	\label{fig:rddf_csites}
\end{figure}

 The COH site also reduced the contribution values for the C site in the case of C$_{60}$(OH)$_{28}$. However, the effect of the polar COH sites was weaker than that of the nonpolar COH site. This means that the order of the effects can be reversed with increasing the number of OH groups. The green bars in Fig. \ref{fig:cvssum} (c) and (d) clearly show the reverse. The reason would also be the differences in the hydration structure. Fig. \ref{fig:rddf_csites} compares the RDDF between the C site and the mass center of water in both the polar (red solid line) and nonpolar (blue solid line) cases for each solute. In C$_{60}$(OH)$_{28}$, the RDDF for the polar case is consistently higher than the nonpolar case until around 0.8 nm (See Fig. \ref{fig:rddf_csites}(c)). However, such behavior is not observed in C$_{60}$(OH)$_{2}$ and C$_{60}$(OH)$_{12}$ (See Fig. \ref{fig:rddf_csites} (a), and (b)). Therefore, the polar C$_{60}$(OH)$_{28}$ would have increased the contribution value of the C site due to increased water density around the C site. The increase in water density would be caused by the charge polarization on the C site, induced by attaching many OH groups. This effect is relatively minor. It could be significant in the case of large graphene molecules, such as nanotubes, because the effects accumulate. By contrast, the effect of charge addition on an OH group is relatively minor compared with that of the roughness caused by an OH group.\par

\begin{figure}[hbt]
	\centering
	\includegraphics[width=\linewidth]{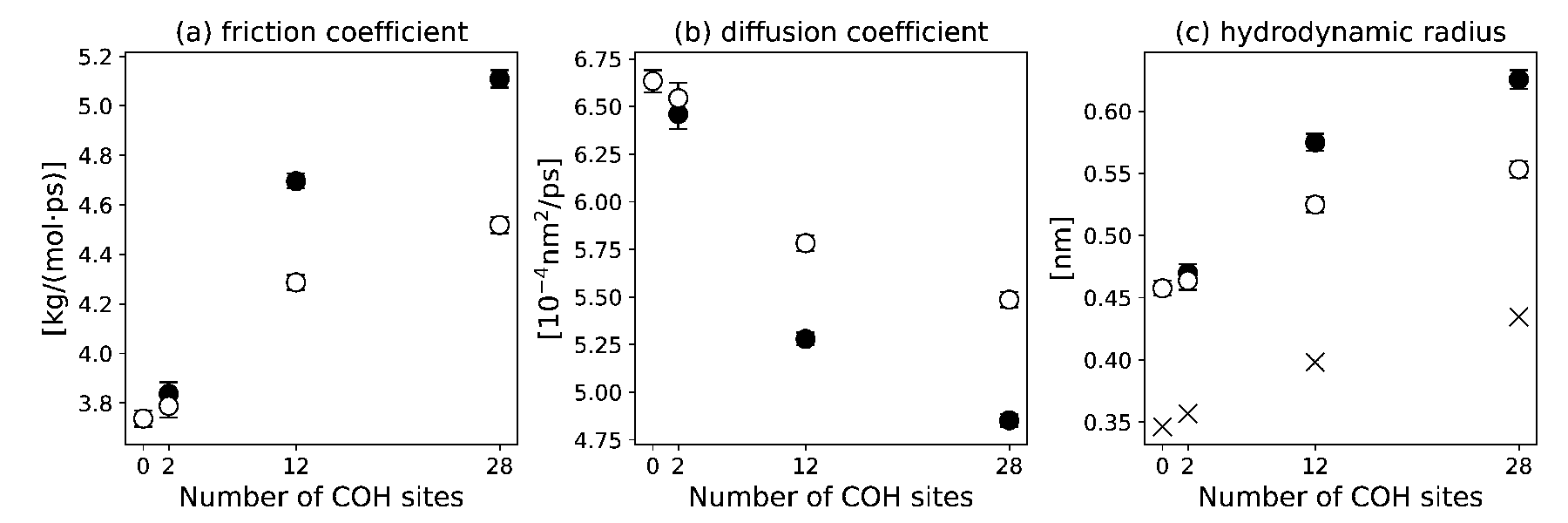}
	\caption{The diffusion properties of fullerene and fullerenols are displayed as a function of the number of COH sites, including (a) friction coefficients, (b) diffusion coefficients, and (c) hydrodynamic radii. The closed and opened circles indicate the results for the polar and nonpolar solute, respectively. In (c), the crosses indicate the gyration radii. The error bars denote the 95\% confidence interval.}
	\label{fig:dps}
\end{figure}

We now discuss the friction coefficients, diffusion coefficients, and hydrodynamic radii for the whole molecule. Fig. \ref{fig:dps} shows the friction coefficients, diffusion coefficients, and hydration radii for fullerene and fullerenols, where closed and open circles indicate the results of polar and nonpolar solutes, respectively. These diffusion properties were corrected using Eq. (\ref{YHeq}). Thus, the values are for the infinite system. The friction coefficients and hydration radii increase with an increasing number of OH groups (See Fig. \ref{fig:dps} (a) and (c)). By contrast, the diffusion coefficients decrease in Fig. \ref{fig:dps} (b). Of course, these dependencies are consistent with each other (See Eqs. (\ref{ESeq}) and (\ref{SEeq})). These differences in diffusion properties are particularly pronounced in the case of polar solutes. For instance, the difference between the polar and the nonpolar C$_{60}$(OH)$_{28}$ is by a factor of about 1.13, where the difference is larger than the errors. These differences are caused by the effect of electrostatic interactions between the molecular surface and solvent water. The charges on the polar sites enhance the local density of the solvent water, and the increases in friction reduce the diffusion coefficients. However, the differences are not only caused by the charges on the surface; the effects of surface roughness are also significant.\par

The fullerenol diffusivity changes with increasing the number of OH groups are qualitatively similar to the previous simulation studies \cite{Chaban,Keshri,Maciel}, in which the fullerenol diffusion coefficients also decrease with increasing number of OH groups. Unfortunately, we did not observe diffusivity changes as large as a factor of 5 to 6, as reported by those studies. This relatively smaller change would originate  from the differences in the MD simulation setup, such as the simulation time length, simulation cell size, force field, thermostats, and barostats. We note that the diffusion coefficients depend on the simulation cell size\cite{YHM} and barostat\cite{NPTdiffusion}, so that we cannot simply compare the values. In the present study, cell-size correction was performed,\cite{YHM} and this is the advantage of our approach.\par

Generally, the relationship between the hydrodynamic and gyration radii depends strongly on the molecular shape, which is characterized by the constant $\rho$ as:
\begin{equation}
R_{\rm g}=\rho \cdot R_{\rm h},
\end{equation}
where, for spherical molecules, the constant $\rho$ is predicted to be $\sqrt{3/5} \approx$ 0.775 \cite{Burchard}. In Fig. \ref{fig:dps} (c), we also plot the gyration radii as crosses where only the radii for polar solutes were plotted because there were no differences with those for nonpolar solutes. The constant $\rho$ exhibits values ranging from 0.693 to 0.785, consistent with the spherical nature of fullerene and fullerenols. Thus, the hydrodynamic radii depend on the molecular radii. In particular, the hydrodynamic radii for nonpolar solutes demonstrate almost identical behaviors to the gyration radii. For instance, the radius ratio between fullerene and the nonpolar C$_{60}$(OH)$_{28}$ is around 1.2 for both the hydrodynamic and gyration radii. By contrast,  the radius ratio between fullerene and the polar C$_{60}$(OH)$_{28}$ is about 1.37 for the hydrodynamic radius (which is larger than the gyration radius). Similar arguments can be applied for C$_{60}$(OH)$_{2}$ and C$_{60}$(OH)$_{12}$. This indicates that the hydrodynamic radius includes the effects of the interaction between the molecular surface and water besides the effect of the molecular radius. Furthermore, from the arguments of contribution values, we can explain in detail that the hydrodynamic radius includes the effects of hydrogen bonds in the case of C$_{60}$(OH)$_{2}$ and C$_{60}$(OH)$_{12}$, as well as the effects of hydrogen bonds and the strong interactions between C sites and water in the case of C$_{60}$(OH)$_{28}$. \par

\begin{figure}[hbt]
	\centering
	\includegraphics[width=\linewidth]{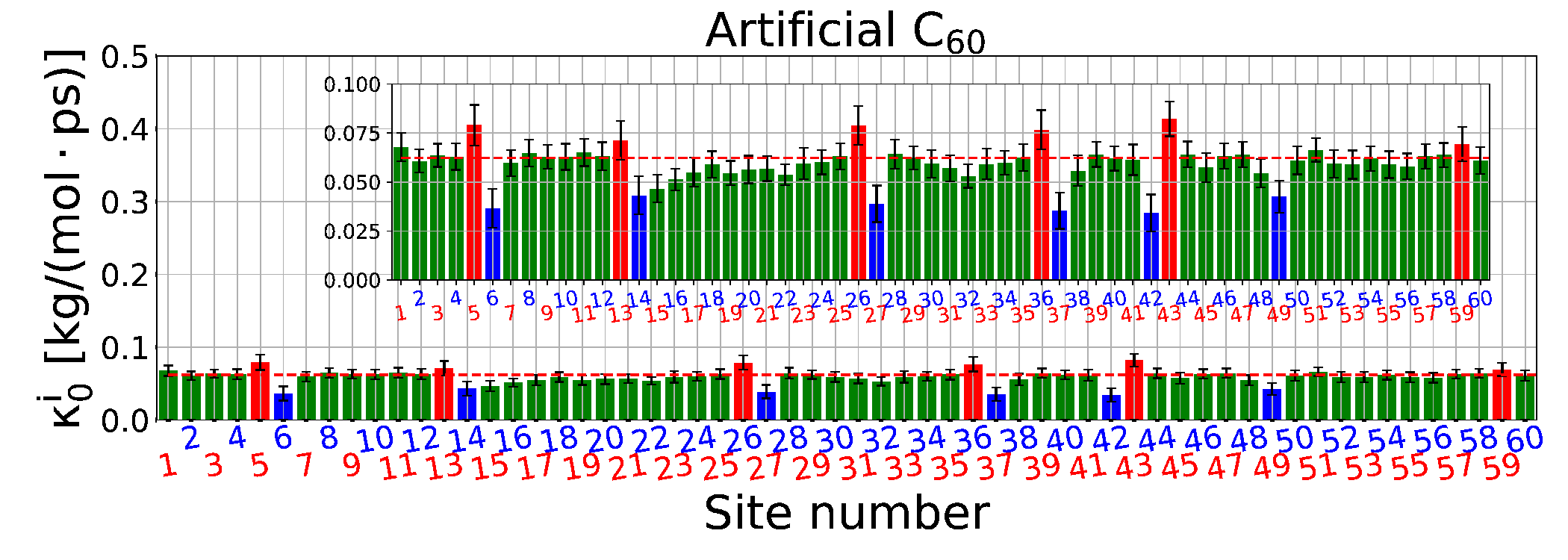}
	\caption{The contribution value of 60 sites on the artificial fullerene model. The red bars (Sites 5, 13, 26, 36, 43, and 59) and the blue bars (Sites 6, 14, 27, 37, 42, and 49) indicate the positively and negatively charged sites, respectively. The other bars (green bars) indicate the normal carbon sites with zero charge. The error bars denote the 95$\%$ confidence interval. The red dashed line indicates the friction coefficient per one C atom on fullerene obtained by dividing the fullerene friction coefficient by 60. In the inset, the range between 0 and 0.1 kg/(mol$\cdot$ ps) is enlarged.}
	\label{fig:art}
\end{figure}

Finally, we report the case of an artificial fullerene model that has charged sites. We placed six positive charges on Sites 5, 13, 26, 36, 43, and 59, and six negative charges on Sites 6, 14, 27, 37, 42, and 49, where site numbers correspond to the number on the fullerene expanded maps (Fig. \ref{fig:cvs}).  The each positively and negatively charged sites exist next to each other and has a charge with an absolute value of 0.3031. We set the charge based on the bonding moment of OH in SPC/E water\cite{SPCE}.\par

Fig.\ref{fig:art} shows the contribution value of 60 sites on the artificial fullerene model. The red bars (Sites 5, 13, 26, 36, 43, and 59) and blue bars (Sites 6, 14, 27, 37, 42, and 49) indicate the positively and negatively charged sites, respectively. The other bars (green bars) indicate the normal C sites with zero charge. In Fig. \ref{fig:art}, the green bars are identical to the friction coefficient per one C atom on fullerene, obtained by dividing the friction coefficient of fullerene by 60  (see red dashed line), while the charged sites are different from the value of the red dashed line. The difference originates from the electrostatic interactions between water and the charged sites. Before the present study, we speculated that the contribution values would increase similarly to those for fullerenol. However, the increase in the contribution value in the positively charged sites was not large. On the contrary, the contribution value in the negatively charged sites was smaller than that of the C site (green bars and the red dashed bar). As a result, the friction coefficient for the artificial fullerene model (3.57 $\pm$ 0.03 kg/(mol $\cdot$ ps)) was smaller than that of fullerene (3.74 $\pm$ 0.03 kg/(mol $\cdot$ ps)). The results showed that our speculation was wrong.\par

The above result shows that the electrostatic interactions between water and the charged site do not always increase the friction. These results can be rationalized. One of the reasons would be a decrease in surface roughness. Because the charges are buried inside the sphere, it was thought that the water could not effectively interact with the charge sites. In particular, the negative charge sites and the positive charge of water are located at the center of the van der Waals sphere, which makes it difficult for them to interact with each other. In other words, the surface roughness, including the surface structures, is also essential for the interaction between the charged sites and water molecules, because the charges cannot increase the accessible area for water. Surprisingly, the friction slightly decreases due to the charges buried on the surface of macromolecules. It seems that the buried OH groups break the hydrogen-bonding network between water molecules and cause a reduction in the local viscosity of water near the macromolecular surface. This effect needs to be clarified in a future study.\par

\section{Conclusion}
The diffusion properties of fullerene and fullerenols  (C$_{60}$(OH)$_{2}$, C$_{60}$(OH)$_{12}$, or C$_{60}$(OH)$_{28}$) were calculated in aqueous environments. Our analysis showed that the diffusion coefficients depend on the surface characteristics of the fullerene C$_{60}$. In particular, the diffusion coefficients of fullerenol with 28 OH groups increased by a factor of about 1.37 over fullerene. This difference is comparable to diffusivity differences accompanied by conformation changes of protein induced by a light reaction observed in the experiments of Terazima and coworkers; for example, the LOV2 -linker is about 1.2 times. Thus, our results support that the minor changes on the macromolecule surface can change the diffusion coefficients significantly. The contribution kernels and values for the specific sites on a molecule were defined using the framework of the GLE. Here, the contribution kernel and value are expected to represent the contribution of a specific site on the friction kernel and coefficient, respectively. We calculated the contribution kernels and values for the specific sites with fullerenols  (C$_{60}$(OH)$_{2}$, C$_{60}$(OH)$_{12}$, or C$_{60}$(OH)$_{28}$) in aqueous environments. A strong correlation was observed between the obtained contribution values and the SASA, which indicates that surface roughness is an important factor for the diffusion coefficient. Large changes in  diffusion were observed due to increasing the number of nonpolar OH in the case of fullerenols with nonpolar OH groups. However, the diffusion changes became larger in the case of fullerenols with polar OH groups. Comparisons between the results of polar and non-polar fullerenols revealed that electrostatic interactions also influence the contribution values. The increased amount of the introduction of the non-polar OH group on the hydrodynamic radius is the same extent as that of the charge addition in the OH group. The charges' effect is comparable with the roughness effect in the present paper. However, the amount of hydrogen bonds between the OH group and the water depends on the level of exposure to the water. Even if the OH group has enough charges, the group buried on the fullerene surface cannot construct the hydrogen bonds with the water, and the effect on the hydrodynamic radius remains only slight. From these results, we conclude that it is important for the macromolecule diffusivities to take the surface roughness into account, as well as the direct interaction between solvent and solute, such as hydrogen bondings.\par

\begin{acknowledgments}
We would like to acknowledge the support of the Kyushu University Leading Human Resources Development Fellowship Program. The computations were performed using the Research Center Computational Science, Okazaki, Japan and the Research Institute for Information Technology, Kyushu University, Japan, respectively.
\end{acknowledgments}


\pagebreak

\begin{center}
\textbf{\large Supplementary information for: Decomposition of Friction Coefficients to Analyze Hydration Effects on a C$_{60}$(OH)$_{\rm n}$}
\end{center}
\setcounter{equation}{0}
\setcounter{section}{0}
\setcounter{figure}{0}
\setcounter{table}{0}
\setcounter{page}{1}
\makeatletter
\renewcommand{\theequation}{S\arabic{equation}}
\renewcommand{\thefigure}{S\arabic{figure}}
\renewcommand{\bibnumfmt}[1]{[S#1]}
\renewcommand{\citenumfont}[1]{S#1}

\section{Belly simulations}
We joined the belly simulations to each other using several quasi-static steps, as described in the following procedure.

\begin{figure}[h]
	\centering
	\includegraphics[width=\linewidth]{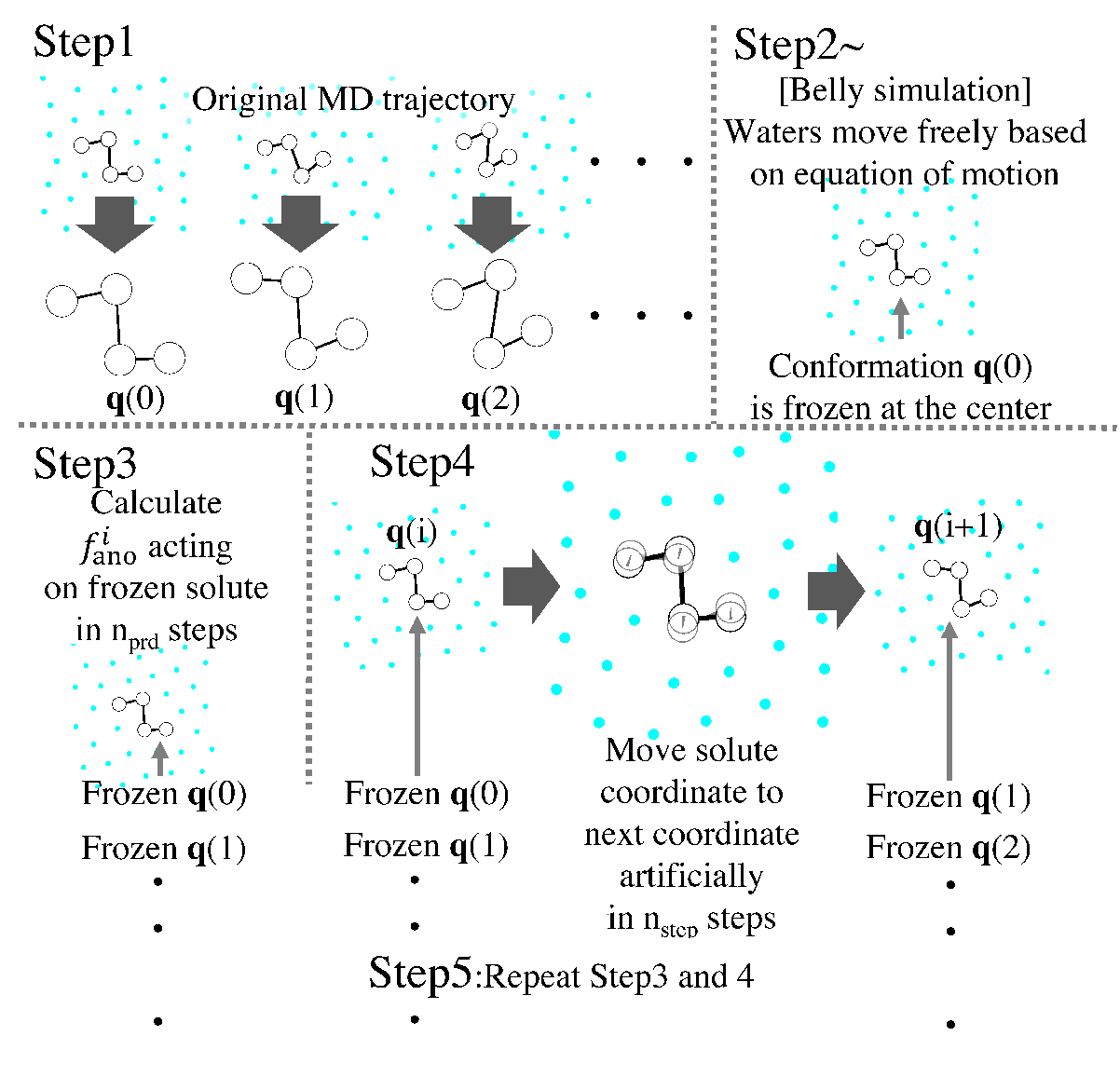}
	\caption{Schematic drawings for steps 1 $\sim$ 5. The linear  four-atomic molecule is the solute, and the small circles denote the solvent molecules.  }
	\label{fig:steps}
\end{figure}

\begin{enumerate}
 \item Acquisition of the target solute conformation: $\lbrack  {\mathbf q_0}(0), {\mathbf q_0}(1)\cdots{\mathbf q_0}(n)\rbrack$ from the original MD trajectory. Here, the original MD trajectory is obtained from the production runs under the normal NVT ensemble.
 
 \item Equilibration of the belly simulation, in which the solute coordinates are frozen at the target solute conformation ${\mathbf q}(0)$.
 
 \item Production run of a belly simulation, in which the solute coordinates are frozen at the target solute conformation, which is ${\mathbf q}(0)$ in the first case, and move the solvent freely over $n_{\rm prd}$ steps. Then, the forces exerting on each atom by the solvents  are averaged to obtain $f^i_{\rm ano}$. To reduce the noise of $f^i_{\rm ano}$, the obtained  forces $f_{\rm ano}^i$ are corrected by the following equation:
\begin{equation}
\label{corr_fano}
\sum_{i=0}^N f_{\rm ano}^i = \sum_{i=0}^N \bigg\lbrack f_{\rm ano}^i - \frac{m^i_{\rm a}}{m}  f_{\rm sum} \bigg\rbrack,
\end{equation}
where $m^i_{\rm a}$ is the mass of the $i$-atom, $f_{\rm sum}$ is the sum of the obtained forces, i.e., $f_{\rm sum}=\sum_{i=0}^N f_{\rm ano, obtained}^i$. The taking average  of $f_{\rm ano}^i$ or Eq. (\ref{corr_fano}) does not change the essential of $f^i_{\rm IN,ano}({\mathbf q}_0)$.

 \item Quasistatic steps, in which the solute coordinates move slowly to the next target solute conformation, which is ${\mathbf q}(1)$ in the first case, over $n_{\rm step}$ steps using the following equation with $j=1,2,...,n_{\rm step}$:
\begin{equation}
{\mathbf X}^j = \frac{(n_{\rm step}-j){\mathbf q}_{\rm pre}+j \cdot {\mathbf q}_{\rm next}}{n_{\rm step}},
\end{equation}
where ${\mathbf X}^j$, ${\mathbf q}_{\rm pre}$, and ${\mathbf q}_{\rm next}$ are the $j$--th solute coordinate, the previous target solute conformation, and the next target solute conformation. The $n_{\rm step}$ is determined from the following equation: 
\begin{equation}
n_{\rm step} = max(disp({\mathbf q}_{\rm next},{\mathbf q}_{\rm pre}))\cdot n_{\rm buffer},
\end{equation}
where $max$ and $disp$ are operators to obtain the maximum value and to obtain the displacement of each atom between the previous and next target solute conformation. This equation ensures that the solute atoms displacement per step is less than 1/$n_{\rm buffer}$ nm. In each step, the solvent molecules move freely.
 \item Repeat steps 3 and 4 to obtain $f_{\rm ano}^i({\mathbf q}(k))$ at the $k$ -th target solute conformation ${\mathbf q}(k)$.
\end{enumerate}
In step 1, to minimize the translational and rotational motion, each structure was superposed upon that of the previous target solute conformation. In step 3, by projecting onto the solute coordinate in the original MD trajectory, the obtained forces can be restored to their original orientation. In step 4, we assumed the quasistatic process. Similar procedures are used in the studies of mean force dynamics (MFD)\cite{MFD1,MFD2,MFD3,MFD4}. The thermodynamic conditions were set the same as the original MD simulation. The correction for the center of mass translational velocity, which was set by default in the GROMACS 2021.4 software, was turned off. We used $n_{\rm prd}=100$ and $n_{\rm buffer}=5000$. To calculate the steps 3 $\sim$ 5, a modified version of GROMACS 2021.4 was prepared \cite{GROMACS_SI}. The code was modified to conduct the steps 3 $\sim$ 5 continuously. The modified code is available at (\url{https://github.com/TomoyaIwashita/gromacs2021_belly}). Fig. \ref{fig:steps} shows the schematic drawings for each step where the solute is expressed by a 2-dimensional linear  four-atomic molecule instead of fullerenols.\par

After obtaining the values  $\lbrace f^i_{\rm ano}({\mathbf q}_0(0)), f^i_{\rm ano}({\mathbf q}_0(1)), \cdots f^i_{\rm ano}({\mathbf q}_0(n))\rbrack$, we can calculate $C^{f(\rm EX,i) v}(t)$ and $C^{f(\rm EX,i) F}(t)$ using Eqs. (16 and (17). In the present study, the interval of the restart points on $(f^i-f^i_{\rm ano})$ was set to 0.1ps.\par

\newpage

\end{document}